\documentclass[runningheads,a4paper]{llncs}

\usepackage{amsmath,amssymb}
\usepackage{graphicx}
\usepackage{caption}
\usepackage{subcaption}
\usepackage{color}
\usepackage{multirow}
\usepackage{url}

\usepackage{mathptmx}
\usepackage{dsfont}
\usepackage{bm}
\usepackage{amstext}

\usepackage{tikz}
\usetikzlibrary{arrows}

\usepackage{array}
\newcolumntype{L}[1]{>{\raggedright\let\newline\\\arraybackslash\hspace{0pt}}m{#1}}
\newcolumntype{C}[1]{>{\centering\let\newline\\\arraybackslash\hspace{0pt}}m{#1}}
\newcolumntype{R}[1]{>{\raggedleft\let\newline\\\arraybackslash\hspace{0pt}}m{#1}}

\newcommand{\E}{\mathbb{E}}
\newcommand{\Var}{\textrm{Var}}
\newcommand{\Cov}{\textrm{Cov}}

\newcommand{\complex}[1]{{\fontfamily{lmss}\selectfont co#1}}
\newcommand{\node}[1]{{\fontfamily{lmss}\selectfont #1}}

\begin{document}

\mainmatter


\title{Reviving the Two-state Markov Chain Approach \\ (Technical Report)}

\titlerunning{Reviving the Two-state Markov Chain Approach}

%

\author{Andrzej Mizera\inst{1} \and Jun Pang\inst{1,2} \and Qixia Yuan\inst{1}\thanks{Supported
by the National Research Fund, Luxembourg (grant 7814267) and partially by Google Summer of Code 2014. }}
\authorrunning{Mizera, Pang and Yuan}

\institute{
University of Luxembourg, FSTC, Luxembourg
\and
University of Luxembourg, SnT, Luxembourg\\
\email{firstname.lastname@uni.lu}
}
\maketitle

\begin{abstract}
Probabilistic Boolean networks (PBNs) is a~well-established computational framework for
modelling biological systems. The steady-state dynamics of PBNs is of crucial importance in
the study of such systems. However, for large PBNs, which often arise in systems biology,
obtaining the steady-state distribution poses a~significant challenge. In fact, statistical
methods for steady-state approximation are the only viable means when dealing with large
networks. In this paper, we revive the two-state Markov chain approach presented in the
literature. We first identify a~problem of generating biased results, due to the size of the
initial sample with which the approach needs to start and we propose a~few heuristics to avoid
such a~pitfall. Second, we conduct an~extensive experimental comparison of the two-state Markov chain
approach and another approach based on the Skart method and we show that statistically the
two-state Markov chain has a~better performance. Finally, we apply this approach to a~large PBN
model of apoptosis in hepatocytes.

\end{abstract}

\section{Introduction}
\label{sec:intro}
Systems biology aims to study biological systems from a~holistic perspective, with the goal to
provide a~comprehensive, system-level understanding of cellular behaviour. Proper functioning of
a~living cell requires a~finely-tuned and orchestrated interplay of many complex processes.
Complex interactions within a~biological system lead to emergent properties which are crucial
for sustaining life. Therefore, understanding the machinery of life requires the use of holistic
approaches which enable the study of a~system as a~whole, in contrast to the reductionist
approach.
Computational modelling plays a~prominent role in the field of systems biology. Construction and
analysis of a~computational model for some biological process enables the systematisation
of available biological knowledge, identification of missing biological information, provides
formal means for understanding and reasoning about the concerted interplay between different parts
of the model, finally reveals directions for future experimental work which could
provide data for better understanding of the process under study.

Unfortunately, computational modelling of biological processes that take place in a~living cell
poses significant challenges with respect to the size of the state-space that needs to be
considered. Modelling of certain parts of cellular machinery such as gene regulatory networks
(GRNs) or signal transduction pathways often leads to dynamical models characterised by huge
state-spaces of sizes that surpass the sizes of any human-designed systems by orders of magnitude.
Therefore, profound understanding of biological processes asks for the development of new methods
and approaches that would provide means for formal analysis and reasoning about such huge systems.

In this study we concentrate on the analysis of the steady-state dynamics of biological processes,
in particular GRNs, modelled as discrete-time Markov chains (DTMCs). This is the case, for
example, when the biological system under study is cast into the mathematical/computational
framework of probabilistic Boolean networks (PBNs)~\cite{SDZ02,TMPTSS13}. In these or other
discrete-time models, e.g., dynamic Bayesian networks, the real (considered as continuous) time
is not modelled. Instead, the evolution of the system is abstracted as a~sequence of consecutive
events. These coarse-grained models have been successfully applied in many systems biology studies
and proved their predictive power~\cite{AO03}. In fact, for the study of large regulatory systems
they remain the only reasonable solution. Extrapolating the ordinary differential equations model
of a~single elementary building block of the network (e.g., a~gene) to the whole large system
would result in a~prohibitively complex model. However, moving towards a~higher-level description
by ignoring the molecular details allows to grasp the system-level behaviour of the
network~\cite{Bornholdt05}.
In consequence, these coarse-grained formalisms are broadly applied in
systems biology studies of systems where the predictions of exact reaction times are not
of main interest. For example, this is the case in the study of dynamical attractors of
a~regulatory network, which seem to depend on the circuit wiring rather than kinetic constants
(such as production rates or interaction rates)~\cite{Wagner05}. In this sense
modelling biological systems with more abstract, high-level view formalisms has certain
unquestionable advantages.

One of the key aspects in the analysis of such dynamic systems is the comprehension of their
steady-state (long-run) behaviour. For example, attractors of such systems were hypothesised to
characterise cellular phenotypes~\cite{Kauffman69a}. 
Another complementary conjecture is that attractors correspond to functional cellular states such
as \emph{proliferation}, \emph{apoptosis}, or \emph{differentiation}~\cite{Huang01}.
These interpretations may cast new light on the understanding of cellular homeostasis and cancer
progression~\cite{SDZ02}. In this work, we focus on the computation of steady-state probabilities
which are crucial for the determination of long-run influences and sensitivities. These are
measures that quantify the impact of genes on other genes, considered collectively or
individually, and that enable the identification of elements with highest impact. In this way they
provide insight into the control mechanisms of the network.

So far the huge-state space, which often characterises dynamical models of biological systems,
tempers the application of the above mentioned techniques in the analysis of realistic biological
systems. In fact, approximations with the use of Markov chain Monte Carlo (MCMC) techniques are
the only viable solution to this problem~\cite{SGHDZ03}. However, due to the difficulties with
the assessment of the convergence rate to the steady-state distribution (see, e.g.,~\cite{CC96}),
certain care is required when applying these methods in practice. A~number of statistical methods
exists, which allow to empirically determine when to stop the simulation and output estimates.
We employ in our study: (1)~the two-state Markov chain approach~\cite{RL92} and (2)~the Skart
batch-means method of~\cite{TWLS08}. The two-state Markov chain approach was introduced in 1992 by
Raftery and Lewis; and Shmulevich et al.~\cite{SGHDZ03} proposed its application to the analysis
of PBNs in 2003. However, to the best of our knowledge, since then it has not been widely applied
for the analysis of large PBNs. In this paper, we aim to revive its usage for approximating
steady-state probabilities of large PBNs, which often arise in computational systems biology as
models of full-size genetic networks. We identify a~problem concerned with the choice of the
initial sample size in the two-state Markov chain approach. As we show, an~unconscious choice may
lead to biased results. We propose a~few heuristics to avoid this problem. By extensive
experiments, we show that the two-state Markov chain approach outperforms the Skart method in most
cases, where the batch-means Skart method is considered the current state-of-the-art approach. In
this way we show that the two-state Markov chain is often the optimal choice for the analysis of
large PBN models of biological systems.

\medskip\noindent
{\bf Structure of the paper.}
After presenting some preliminaries in Section~\ref{sec:pre}, we describe the two-state Markov
chain approach in Section~\ref{sec:twostate} and identify a~problem of generating biased results,
due to the size of the initial sample with which the approach needs to start. We then propose some
heuristics for the approach to avoid unfortunate initialisations. We perform an~extensive
evaluation and comparison of the two-state Markov chain approach and the Skart method in
Section~\ref{sec:evaluation} on a~large number of randomly generated PBNs. In most cases, the
two-state Markov chain approach seems to have a~better performance in terms of computational cost.
Finally, we apply the two-state Markov chain approach to study a~large PBN model of apoptosis in
hepatocytes consisting of $91$ nodes in Section~\ref{sec:casestudy}. We compute
the steady-state influences and long-run sensitivities and confirm previously formulated
hypothesis.
We conclude our paper with some discussions in Section~\ref{sec:conclusion}.

\section{Preliminaries}
\label{sec:pre}

\subsection{Finite discrete-time Markov chains (DTMCs)}
\label{ssec:dtmc}
Let $S$ be a~finite set of states. A~(first-order) discrete-time Markov chain is an~$S$-valued
stochastic process $\{X_t\}_{t\in \mathbb{N}}$ with the property that the next state is
independent of the past states given the present state. Formally,
$\mathbb{P}(X_{t+1}=s_{t+1}\,|\,X_t = s_t,X_{t-1} = s_{t-1},\ldots,X_0=s_0) =
\mathbb{P}(X_{t+1}=s_{t+1}\,|\,X_t = s_t)$ for all $s_{t+1},s_t,\ldots, s_0 \in S$. Here, we
consider \emph{time-homogenous} Markov chains, i.e., chains where
$\mathbb{P}(X_{t+1}=s'\,|\,X_t=s)$, denoted $P_{s,s'}$, is independent of $t$ for any states
$s,s'\in S$. The transition matrix $P=(P_{s,s'})_{s,s'\in S}$ satisfies
$P_{s,s'}\geqslant 0$ and $\sum_{s'\in S}P_{s,s'}=1$ for all $s\in S$.
We denote by $\pi$ a~probability distribution on $S$. If $\pi = \pi\, P$, then $\pi$ is
a~\emph{stationary distribution} of the DTMC (also referred to as the \emph{invariant
distribution}). A~path of length $n$ is a~sequence $s_1\to s_2\to\cdots\to s_{n}$ such that
$P_{s_{i},s_{i+1}}>0$ and $s_i \in S$ for $i \in \{1,2,\ldots, n\}$. State $q\in S$ is
\emph{reachable} from state $p\in S$ if there exists a~path such that $s_1 = p$ and $s_{n} = q$.
A~DTMC is \emph{irreducible} if any two states are reachable from each other. The period of
a~state is defined as the greatest common divisor of the lengths of all paths that start and end
in the state. A~DTMC is \emph{aperiodic} if all states in $S$ are of period $1$. A~finite state
DTMC is called \emph{ergodic} if it is irreducible and aperiodic. By the famous ergodic theorem
for DTMCs~\cite{Norris98} an~ergodic chain has a~unique stationary distribution being its
\emph{limiting distribution} (also referred to as the \emph{steady-state distribution} given by
$\lim_{n\to\infty}\pi_0\,P^n$, where $\pi_0$ is any initial probability distribution on $S$). In
consequence, the limiting distribution for an~ergodic chain is independent of the choice of
$\pi_0$. The steady-state distribution can be estimated from any initial distribution by
iteratively multiplying it by $P$.

The evolution of a~first-order DTMC can be described by a~stochastic recurrence sequence $X_{t+1}
= \phi(X_t,U_{t+1})$, where $\{U_{t}\}_{t\in \mathbb{N}}$ is an~independent sequence of uniformly
distributed real random variables over $[0,1]$ and the transition function $\phi:S \times [0,1]
\to S$ satisfies the property that $\mathbb{P}(\phi(s,U)=s')=P_{s,s'}$ for any states $s,s' \in
S$ and for any $U$, a~real random variable uniformly distributed over $[0,1]$. When $S$ is
partially ordered and when the transition function $\phi(\cdot,u)$ is monotonic for all $u$, then
the Markov chain is said to be \emph{monotone} (\cite{PW96,BGV08}).

\subsection{Probabilistic Boolean networks (PBNs)}
\label{ssec:pbn}
A~PBN $G(V,\mathcal{F})$ consists of a~set of binary-valued nodes (also known as genes)
$V=\{v_{1},v_{2}, \ldots ,v_{n}\}$ and a~list of sets $\mathcal{F}=(F_{1},F_{2}, \ldots, F_{n})$.
For each $i \in \{1,2,\ldots,n\}$ the set
$F_{i}=\{f_{1}^{(i)},f_{2}^{(i)},\ldots,f_{l(i)}^{(i)}\}$ is a~collection of predictor functions
for node $v_{i}$, where $l(i)$ is the number of predictor functions for $v_{i}$. Each $f_{j}^{(i)}
\in F_{i}$ is a~Boolean function defined with respect to a~subset of nodes referred to as parent
nodes of $v_i$. There is a~probability distribution on each $F_i \in F$: $c_{j}^{(i)}$ is the
probability of selecting $f_j^{(i)} \in F_i$ as the next predictor for $v_i$ and it holds that
$\sum_{j=1}^{l(i)}c_{j}^{(i)}=1$. We denote by $v_{i}(t)$ the value of node $v_{i}$ at time point
$t\in \mathbb{N}$. The state space of the PBN is $S=\{0,1\}^n$ and it is of size $2^n$. The state
of the PBN at time $t$ is given by $\boldsymbol{s}(t)=(v_{1}(t),v_{2}(t),\ldots,v_{n}(t))$. The
dynamics of the PBN is given by the sequence $(\boldsymbol{s}(t))_{t=0}^\infty$. We consider here
\emph{independent} PBNs where predictor functions for different nodes are selected independently
of each other. The transition from $\boldsymbol{s}(t)$ to $\boldsymbol{s}(t+1)$ is conducted by
randomly selecting a~predictor function for each node $v_i$ from $F_i$ and by synchronously
updating the node values in accordance with the selected functions. There are
$N=\prod_{i=1}^{n}l(i)$ different ways in which the predictors can be selected for all $n$ nodes.
These combinations are referred to as \emph{realisations} of the PBN and are represented as
$n$-dimensional function vectors $\boldsymbol{f}_k =
(f_{k_1}^{(1)},f_{k_2}^{(2)},\ldots,f_{k_n}^{(n)})\in F_1\times F_2\times \ldots \times F_n$,
where $k\in \{1,2,\ldots,N\}$ and $k_i \in \{1,2,\ldots,l(i)\}$. A~realization selected at time
$t$ is referred to as $\boldsymbol{f}(t)$. Due to independence, $\mathbb{P}(\boldsymbol{f}_k) =
\mathbb{P}(\boldsymbol{f}(t) = \boldsymbol{f}_k) = \prod_{i=1}^n c_{k_i}^{(i)}$.

In PBNs \emph{with perturbations}, a~perturbation parameter $p \in (0,1)$ is introduced to sample
the perturbation vector
$\boldsymbol{\gamma}(t)=(\gamma_{1}(t),\gamma_{2}(t),\ldots,\gamma_{n}(t))$, where $\gamma_{i}(t)
\in \{0,1\}$ and $\mathbb{P}(\gamma_{i}(t)=1)=p$ for all $t$ and $i \in \{1,2,\ldots,n\}$.
Perturbations provide an~alternative way to regulate the dynamics of a~PBN: the next state is
determined as $\boldsymbol{s}(t+1)= \boldsymbol{f}(t)(\boldsymbol{s}(t))$ if
$\boldsymbol{\gamma}(t)=\boldsymbol{0}$ and as $\boldsymbol{s}(t+1)=\boldsymbol{s}(t) \oplus
\boldsymbol{\gamma}(t)$ otherwise, where $\oplus$ is the exclusive or operator for vectors. The
perturbations, by the latter update formula, allow the system to move from any state to any other
state in one single transition, hence render the underlying Markov chain irreducible and
aperiodic. Therefore, the dynamics of a~PBN with perturbations can be viewed as an~ergodic
DTMC~\cite{SD10}. The transition matrix is given by
$P_{s,s'}=(1-p)^n \sum_{k=1}^{N}\boldsymbol{1}_{[\boldsymbol{f}_k(s)=s']}
\mathbb{P}(\boldsymbol{f}_k)+(1-(1-p)^n)p^{\eta(s,s')}(1-p)^{n-\eta(s,s')}$, where
$\boldsymbol{1}$ is the indicator function and $\eta(s,s')$ is the hamming distance between states
$s,s' \in S$.
According to the ergodic theory, adding perturbations to any PBN assures the long-run dynamics of
the resulting PBN is governed by a~unique limiting distribution, convergence to which is
independent of the choice of the initial state. However, the perturbation probability value should
be chosen carefully, not to dilute the behaviour of the original PBN. In this way the
`mathematical trick', although introduces some noise to the original system, allows to
significantly simplify the analysis of the steady-state behaviour, which is often of interest for
biological systems.

The density of a~PBN is measured with its function number and parent nodes number. For a~PBN $G$,
its density is defined as $\mathcal{D}(G)=\frac{1}{n}\sum_{i=1}^{\it N_F}\omega(i)$, where $n$ is
the number of nodes in $G$, ${\it N_F}$ is the total number of predictor functions in $G$, and
$\omega(i)$ is the number of parent nodes for the $i$th predictor function.

Within the framework of PBNs the concept of influences is defined; it formalizes the impact
of parents nodes on a~target node and enables its quantification (\cite{SDKZ02}). The concept is
based on the notion of a~partial derivative of a~Boolean function $f$ with respect to variable
$x_j$ ($1 \leq j \leq n$):
\begin{equation*}
\frac{\partial f(x)}{\partial x_j} =  f(x^{(j,0)}) \oplus f(x^{(j,1)}),
\end{equation*}
where $\oplus$ is addition modulo $2$ (exclusive OR) and for $l \in \{0,1\}$
\begin{equation*}
x^{(j,l)} = (x_1,x_2,\ldots,x_{j-1},l,x_{j+1},\ldots,x_n).
\end{equation*}
The \emph{influence of node $x_j$ on function $f$} is the expected value of the partial derivative
with respect to the probability distribution $D(x)$:
\begin{equation*}
I_j(f) = \mathbb{E}_D\Big[\frac{\partial f(x)}{\partial x_j}\Big] =
 \mathbb{P}\Big\{\frac{\partial f(x)}{\partial x_j} = 1\Big\} =
 \mathbb{P}\{f(x^{(j,0)}) \neq f(x^{(j,1)})\}.
\end{equation*}
Let now $F_i$ be the set of predictors for $x_i$ with corresponding probabilities $c_j^{(i)}$ for
$j=1,\ldots,l(i)$ and let $I_k(f_j^{(i)})$ be the influence of node $x_k$ on the predictor
function $f_j^{(i)}$. Then, the \emph{influence of node $x_k$ on node $x_i$} is
defined as:
\begin{equation*}
I_k(x_i) = \sum_{j=1}^{l(i)}I_k(f_j^{(i)})\cdot c_j^{(i)}.
\end{equation*}
The \emph{long-term influences} are the influences computed when the distribution $D(x)$ is the
stead-state distribution of the PBN.

We define and consider in this study two types of long-run sensitivities.

\begin{definition}
\label{def:LRS_prob}
The \emph{long-run sensitivity with respect to selection probability perturbation} is defined as
\begin{equation*}
s_c[c_j^{(i)}=p] = \|\mathbb{\tilde{\pi}}[c_j^{(i)}=p] - \mathbb{\pi} \|_l,
\end{equation*}
where $\|\cdot\|_l$ denotes the $l$-norm, $\mathbb{\pi}$ is the steady-state distribution of the
original PBN, $p \in [0,1]$ is the new value for $c_j^{(i)}$, and
$\mathbb{\tilde{\pi}}[c_j^{(i)}=p]$ is the steady-state probability distribution of the PBN
perturbed as follows. The $j$th selection probability for node $x_i$ is replaced with
$\tilde{c}_j^{(i)} = p$ and all $c_k^{(i)}$ selection probabilities for $k \in I_{-j} =
\{1,2,\ldots,j-1,j+1,\dots,l(i)\}$ are replaced with
\begin{equation*}
\tilde{c}_k^{(i)} = c_k^{(i)} + (c_j^{(i)}-p)\cdot\frac{c_k^{(i)}}{\sum_{l\in I_{-j}} c_l^{(i)}},
\end{equation*}
The remaining selection probabilities of the original PBN are unchanged.
\end{definition}

\begin{definition}
\label{def:LRS_node}
The \emph{long-run sensitivity with respect to permanent on/off perturbations of a~node $x_i$} as
\begin{equation*}
s_g[x_i] = \max \{\|\mathbb{\tilde{\pi}}[x_i \equiv 0] - \mathbb{\pi} \|_l,
 \| \mathbb{\tilde{\pi}}[x_i \equiv 1] - \mathbb{\pi} \|_l\},
\end{equation*}
where $\mathbb{\pi}$, $\mathbb{\tilde{\pi}}[x_i \equiv 0]$, and $\mathbb{\tilde{\pi}}[x_i \equiv
1]$ are the steady-state probability distributions of the original PBN, of the original PBN with
all $f^{(i)} \in F_{i}$ replaced by $\tilde{f}^{(i)} \equiv 0$, and all $f^{(i)} \in F_{i}$
replaced by $\tilde{f}^{(i)} \equiv 1$, respectively.
\end{definition}

Notice that the definition of long-run sensitivity with respect to permanent on/off perturbations
is similar but not equivalent to the definition of long-run sensitivity with respect to 1-gene
function perturbation of~\cite{SDKZ02}.

\section{The Two-state Markov Chain Approach}
\label{sec:twostate}

\subsection{Description}
\begin{figure}[t]
  \centering
  \begin{subfigure}[b]{0.48\textwidth}
    \includegraphics[scale=0.35]{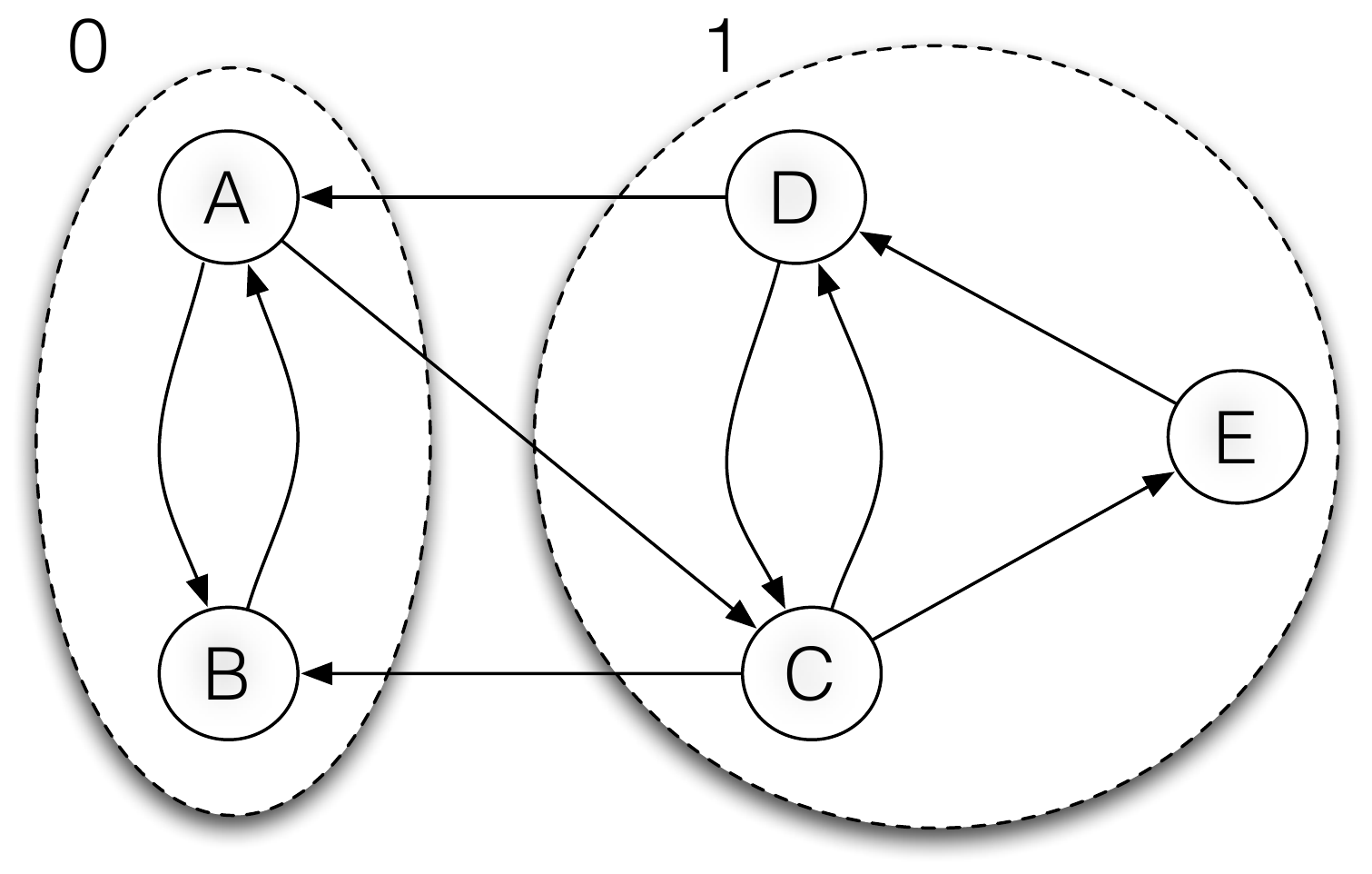}
    \caption{Original DTMC}
    \label{fig:dtmc}
  \end{subfigure}%
  \quad
  \begin{subfigure}[b]{0.48\textwidth}
    \includegraphics[scale=0.34]{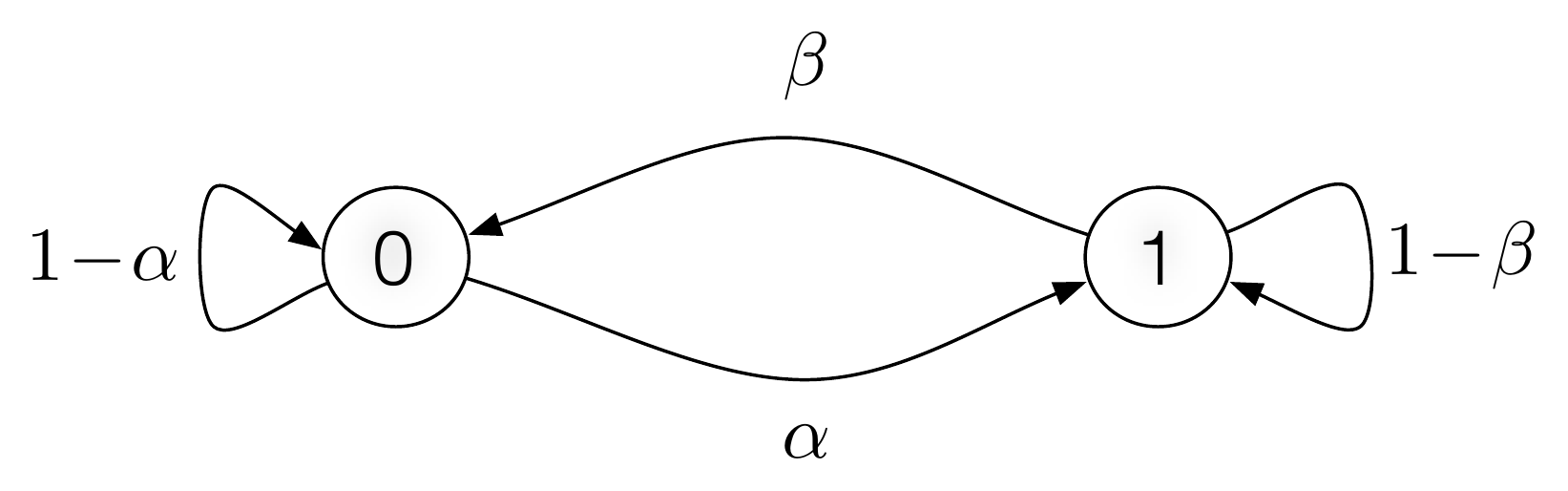}
    \vspace{5mm}
    \caption{Two-state DTMC }
    \label{fig:two-state_MC}
  \end{subfigure}
  \caption{Conceptual illustration of the idea of the two-state Markov chain construction. (a) The
  state space of the original discrete-time Markov chain is split into two meta states: states
  $A$ and $B$ form meta state~$0$, while states $D$, $C$, and $E$ form meta state~$1$. The split
  of the state space into meta states is marked with dashed ellipses. (b) Projecting the behaviour
  of the original chain on the two meta states results in a~binary (0-1) stochastic process. After
  potential subsampling, it can be approximated as a~first-order, two-state Markov chain with the
  transition probabilities $\alpha$ and $\beta$ set appropriately.}
  \label{fig:two-state_MC_approach}
\end{figure}

We recall the two-state Markov chain approach originally introduced in~\cite{RL92}. The two-state
Markov chain approach is a~method for estimating the steady-state probability of a~subset of
states of a~DTMC. In this approach the state space of an~arbitrary DTMC is split into two disjoint
sets, referred to as meta states. One of the meta states, numbered $1$, is the subset of interest
and the other, numbered $0$, is its complement. The steady-state probability of meta state $1$,
denoted $q$, can be estimated by performing simulations of the original Markov chain. For this
purpose a~two-state Markov chain abstraction of the original DTMC is considered. Let
$\{Z_t\}_{t\geqslant 0}$ be a~family of binary random variables, where $Z_t$ is the number of the
meta state the original Markov chain is in at time $t$. $\{Z_t\}_{t\geqslant 0}$ is a~binary (0-1)
stochastic process, but in general it is not a~Markov chain. However, as argued in~\cite{RL92},
a~reasonable assumption is that the dependency in $\{Z_t\}_{t\geqslant 0}$ falls off rapidly with
lag. Therefore, a~new process $\{Z_t^{(k)}\}_{t\geqslant 0}$, where $Z_t^{(k)}=Z_{1+(t-1)k}$, will
be approximately a~first-order Markov chain for $k$ large enough. A~procedure for determining
appropriate $k$ is given in~\cite{RL92}. The first-order Markov chain consists of the two meta
states with transition probabilities $\alpha$ and $\beta$ between them. See
Figure~\ref{fig:two-state_MC_approach} for a~conceptual illustration of the construction of this
abstraction.


The steady-state probability estimate $\hat{q}$ is computed from a~simulated trajectory of the
original DTMC. The key point is to determine the optimal length of the trajectory. Two
requirements are imposed. First, the abstraction of the DTMC, i.e., the two-state Markov chain,
should converge close to its steady-state distribution $\pi = [\pi_0\ \pi_1]$. Formally, $t$
satisfying $|\mathbb{P}[Z_t^{(k)}=i\,|\,Z_0^{(k)}=j]-\pi_i|<\epsilon$ for a~given $\epsilon>0$ and
all $i,j \in \{0,1\}$ needs to be determined. $t$ is the so-called `burn-in' period and determines
the part of the trajectory of the two-state Markov chain that needs to be discarded. Second, the
estimate $\hat{q}$ is required to satisfy $\mathbb{P}[q-r\leqslant\hat{q}\leqslant q+r]\geqslant
s$, where $r$ is the required precision and $s$ is a~specified confidence level. This condition is
used to determine the length of the second part of the trajectory used to compute $\hat{q}$, i.e.,
the sample size. Now, the total required trajectory length of the original DTMC is then given by
$M+N$, where $M=1+(t-1)k$ and $N=1+(\lceil n(\alpha,\beta)\rceil - 1)k$, where $t=\lceil
m(\alpha,\beta)\rceil$. The functions $m$ and $n$ depend on the transitions probabilities $\alpha$
and $\beta$ and are given by
\begin{equation*}
m(\alpha,\beta)=\frac{\log{\left(\frac{\epsilon(\alpha+\beta)}{\max(\alpha,\beta)}\right)}}
{\log{(|1-\alpha-\beta|)}}
\ \textrm{and}\
n(\alpha,\beta)=\frac{\alpha\beta(2-\alpha-\beta)}{(\alpha+\beta)^3}
\frac{\left(\mathrm{\Phi}^{-1}(\frac{1}{2}(1+s))\right)^2}{r^2},
\end{equation*}
where $\mathrm{\Phi}^{-1}$ is the inverse of the standard normal cumulative distribution function.
The expressions for $m$ and $n$ were originally presented in~\cite{RL92}. Derivations however were
not provided and the expressions contain two oversights: in the formula for $m$ the absolute value
is missing in the denominator and in the formula for $n$ the inverse of $\mathrm{\Phi}$ should be
used instead of $\mathrm{\Phi}$. We provide detailed derivations of the expressions for $m$ and
$n$ in the Appendices~\ref{app:m} and \ref{app:n}, respectively.


Since $\alpha$ and $\beta$ are unknown, they need to be estimated. This is achieved iteratively in
the two-state Markov chain approach of~\cite{RL92}. It starts with sampling an~arbitrary initial
length trajectory, which is then used for estimating the values of $\alpha$ and $\beta$. $M$ and
$N$ are calculated based on these estimates. Next, the trajectory is extended to reach the
required length, and $\alpha$ and $\beta$ values are re-estimated. The new estimates are used to
re-calculate $M$ and $N$. This process is iterated until $M+N$ is smaller than the current
trajectory length. Finally, the resulting trajectory is used to estimate the steady-state
probability of meta state~$1$. For more details, see~\cite{RL92}.

The two-state Markov chain approach can be viewed as an~aggregation method for reducing the state
space. Generically, aggregation methods aggregate groups of nodes in the original Markov chain in
accordance with a~given partition function, which leads to a~smaller transition graph. In
principle, the aggregation can be any Markov chain over this smaller transition system. What
remains is the choice of the specific aggregation which is determined by the choice of the
transition probabilities. The partition function is also used to obtain the so-called
\emph{projection} of the original process, i.e., the realisation of the original Markov chain is
projected through the partition function. Ideally, the aggregated chain and the projected
realisation should coincide. However, since the projection is in general not Markovian, the
aggregation which is ``closest'' to the projection is considered instead and the ``closeness'' has
to be appropriately defined. The problem of finding the optimal partition function in the case
where the distance between the projection and the aggregation is quantified by the
Kullback-Leibler divergence rate (KLDR) has been recently studied, see,
e.g.,~\cite{DMM09,DMM11,GPKK15}. The focus in these works is on finding the optimal partition
function, for which the KLDR distance between the projection and aggregation is minimised.
In our case the partition function which classifies the original states into the two meta states
is specified by the biological question under study. As shown in~\cite{DMM11} and \cite{GPKK15},
for a~given partition function, the aggregation closest to the projection can be analytically
obtained provided the steady-state distribution of the original chain is available. The
steady-state probabilities are however our goal, thus these techniques cannot be exploited for the
determination of $\alpha$ and $\beta$ transition probabilities. The iterative, statistical
estimation of the transition probabilities for the aggregation remains the only viable
solution.

\subsection{The choice of the initial sample size}
Given good estimates of $\alpha$ and $\beta$, the theory of the two-state Markov chain presented
above guarantees that the obtained value satisfies the imposed precision requirements.
However, the two-state Markov chain approach starts with generating a~trajectory of the original
DTMC of an~arbitrarily chosen initial length, i.e., $M_0 + N_0 = 1+(m_0-1)k + 1+(n_0-1)k$, where
$m_0$ it the `burn-in' period and $n_0$ is the sample size of the two-state Markov chain
abstraction. An~unfortunate choice may lead to first estimates of $\alpha$ and $\beta$ that are
biased and result in the new values of $M$ and $N$ such that $M+N$ is either smaller or not much
larger than the initial $M_0+N_0$. In the former case the algorithm stops immediately with the
biased values for $\alpha$, $\beta$ and, more importantly, with an~estimate for the steady-state
probability that does not satisfy the precision requirements. The second case may lead to the same
problem. As an~example we considered a~two-state Markov chain with $\alpha=\frac{24}{11873}$
($0.0020214$) and $\beta=\frac{24}{25}$ ($0.96$). The steady-state probability distribution was
$[0.997899\ 0.002101]$. With $k=1$, $\epsilon = 10^{-6}$, $r = 10^{-3}$, $s = 0.95$, $m_0 = 5$,
and $n_0 = 1,920$ the first estimated values for $\alpha$ and $\beta$ were $\frac{1}{1918}$
($0.0005214$) and $1$, respectively. This subsequently led to $M=2$ and $N=1,999$, resulting in
a~request for the extension of the trajectory by $76$. After the extension, the new estimates for
$\alpha$ and $\beta$ were $\frac{1}{1997}$ and $1$, respectively. These estimates gave $M=2$,
$N=1,920$, and the algorithm stopped. The estimated steady-state probability distribution was
$[0.99950\ 0.00050]$, which was outside the pre-specified precision interval given by $r$.
Independently repeating the estimation for $10^{4}$ times resulted in estimates of the
steady-state probabilities that were outside the pre-specified precision interval $90\%$ of times.
Given the rather large number of repetitions, it can be concluded that the specified $95\%$
confidence interval was not reached in this case.

The reason for the biased result is the unfortunate initial value for $n_0$ and the fact that
the real value of $\alpha$ is small. In the initialisation phase the value of $\alpha$
is underestimated and $\lceil n(\alpha,\beta)\rceil$ calculated based on the estimated values of
$\alpha$ and $\beta$ is almost the same as $n_0$. Hence, subsequent extension of the trajectory
does not provide any improvement to the underestimated value of $\alpha$ since the elongation is
short and the algorithm halts after the next iteration.

To identify and avoid some of such pitfalls, we consider a~number of cases and formulate some
of the conditions in which the algorithm may fail to achieve the specified precision.
Let $n_0$ be the initial size of the sample used for initial estimation of $\alpha$ and $\beta$.
We assume that neither $\alpha$ nor $\beta$ is zero. Then, the smallest possible estimates for
both $\alpha$ and $\beta$ are greater than $\frac{1}{n_0} $. Let us set an~upper bound value for
$n_0$ to be $10^4$. For most cases this boundary value is reasonable and we expect $n_0$ to be
much smaller. Notice however that this is the case only if the real values of $\alpha$ and $\beta$
are larger than $10^{-4}$. We just mention here that in general the selection of a~proper value
for $n_0$ heavily depends on the real values of $\alpha$ and $\beta$, which are unknown
\emph{a~priori}. From what was stated above, it follows that both first estimates for $\alpha$ and
$\beta$ are greater than $10^{-4}$. The following cases are possible.
\begin{itemize}
  \item If both $\alpha$ and $\beta$ are small, e.g., less than $0.1$, then we have that
    $10^{-4}<\alpha,\beta<0.1$ and $n(\alpha,\beta) > 72,765$ as can be seen by investigating the
    $n(\cdot,\cdot)$ function. In this case the sample size is increased more than $7$-fold
    which is reasonable since the two-state Markov chain seems to be bad-mixing by the first
    estimates of the values for $\alpha$ and $\beta$ and the algorithm asks for a~significant
    increase of the sample size. We therefore conclude that the bad-mixing Markov chain case can
    be properly handled by the algorithm.

  \item Both first estimates of $\alpha$ and $\beta$ are close to $1$. If $\alpha,\beta \in
    [0.7,0.98]$, the value of $n(\alpha,\beta)$ is larger than $19,000$. If both $\alpha,\beta >
    0.98$ than the size of the sample drops, but in this case the Markov chain is highly
    well-mixing and short trajectories are expected to provide good estimates.

  \item The situation is somewhat different if one of the parameters is estimated to be small
    and the other is close to $1$ as in the example described above. The extension to the
    trajectory is too small to significantly change the estimated value of the small parameter
    and the algorithm halts.
\end{itemize}

\noindent
Considering the above cases leads us to the observation that the following situation needs to be
treated with care:
  \emph{The estimated value for one of the parameters is close to $\frac{1}{n_0}$, the value of
  the second parameter is close to $1$, and $n(\alpha,\beta)$ is either smaller or not
  significantly larger than $n_0$.}


\medskip
\noindent
{\bf First approach: pitfall avoidance.}
In order to avoid this situation, we determine $n_0$ which in principle could lead to initial
inaccurate estimates of $\alpha$ or $\beta$ and such that the next sample size given by $\lceil
n(\alpha,\beta)\rceil$ would practically not allow for an~improvement of the estimates. We reason
as follows. As stated above, the `critical' situation may take place when one of the parameters is
estimated to be very small, i.e., close to $\frac{1}{n_0}$, and the increase in the sample size is
not significant enough to improve the estimate. If the initial estimate is very small, the real
value is most probably also small, but the estimate is not accurate. If the value is
underestimated to the lowest possible value, i.e., $\frac{1}{n_0}$, on average the improvement can
take place only if the sample size is increased at least by $n_0$. Therefore, with the trade-off
between the accuracy and efficiency of the method in mind, we propose the sample size to be
increased at least by $n_0$. Therefore, the `critical' situation condition is
$n(\alpha,\beta)<2\,n_0$. By analysing the function $n(\cdot,\cdot)$ as described in details in
Appendix~\ref{app:1st_approach}, we can determine the values of $n_0$ that are `safe', i.e.,
which do not satisfy the `critical' condition. We present them in Table~\ref{tab:n_0} for a~number
of values for $r$ and $s$.


\begin{table}[t]
  \centering
  \begin{tabular}{|c|c|c|c|c|c|c|c|c|c|}
  \hline
  $r$ & \multicolumn{3}{c|}{$0.01$} & \multicolumn{3}{c|}{$0.001$} &
   \multicolumn{3}{c|}{$0.0001$}\\
  \hline
  $s$ & $0.9$ & $0.95$ & $0.975$ & $0.9$ & $0.95$ & $0.975$ & $0.9$ & $0.95$ & $0.975$\\
  \hline
  $n_0 \in$ & $\emptyset$ & $[2,136]$ & $\emptyset$ & $[2,1161]$ & $[2,1383]$ & $[2,1582]$ &
   $[2,11628]$ & $[2,13857]$ & $[2,15847]$\\
  \hline
  \end{tabular}
  \caption{Ranges of integer values for $n_0$ that
  do not satisfy the `critical' condition $n(\alpha,\beta)<2\,n_0$
  for the given values of $r$ and $s$.}\label{tab:n_0}
\end{table}


\medskip
\noindent
{\bf Second approach: controlled initial estimation of $\alpha$ and $\beta$.}
The formula for $n$ is asymptotically valid for a~two-state Markov chain provided that the values
for $\alpha$ and $\beta$ are known. However, these values are not known \emph{a priori} and they
need to be estimated. Unfortunately, the original approach does not provide any control over the
quality of the initial estimate of the values of these parameters. In certain situation, e.g., as
in the case discussed above, the lack of such a~control mechanism may lead to results with worse
statistical confidence level than the specified one given by $s$. In the discussed example
$s=95\%$, but this value was not reached in the performed experiment. In order to address this
problem, we propose to extend the initial phase of the two-state approach algorithm in the
following way. The algorithm samples a~trajectory of the given Markov chain and estimates the
values of $\alpha$ and $\beta$. It might be the case that an~arbitrarily chosen initial sample
size is not big enough to provide non-zero estimates for the two parameters. If this is the case,
the initial sample size is doubled and the trajectory is elongated to collect a~sample of required
size. This is repeated iteratively until non-zero estimates for both $\alpha$ and $\beta$ are
obtained. We introduce the following notation: $\hat{\alpha}$ and $\hat{\beta}$ are the non-zero
estimates of the values of $\alpha$ and $\beta$, respectively. Furthermore, let $n_0$ be the
sample size used to obtain first non-zero estimates $\hat{\alpha}$ and $\hat{\beta}$, i.e., $n_0$
is either the initial sample size or it is two to power some multiple of the initial sample size.

Once the non-zero estimates are available, the algorithm computes the sample size required to
reach the $s$ confidence level that the true value of $\min(\alpha,\beta)$ is within
a~certain interval. For definiteness, let us assume from now on that $\hat{\alpha}<\hat{\beta}$,
which suggests that $\min(\alpha,\beta)=\alpha$. During the execution of the procedure outlined in
the following the inequality may be inverted. If this is the case, the algorithm makes 
corresponding change in the consideration of $\alpha$ and $\beta$.

The aim is to have a~good estimate for $\alpha$. Notice that the smallest possible initial value
of $\hat{\alpha}$ is greater than $\frac{1}{n_0}$. We refer to $\frac{1}{n_0}$ as the
\emph{resolution of estimation}. Given the resolution, one cannot distinguish between values of
$\alpha$ in the interval $(\hat{\alpha}-\frac{1}{n_0},\hat{\alpha}+\frac{1}{n_0})$. In
consequence, if $\alpha \in (\hat{\alpha}-\frac{1}{n_0},\hat{\alpha}+\frac{1}{n_0})$, then the
estimated value $\hat{\alpha}$ should be considered as optimal. Hence, one could use this interval
as the one which should contain the real value with specified confidence level. Nevertheless,
although the choice of this interval usually leads to very good results, as experimentally
verified, the results are obtained at the cost of large samples which make the algorithm stop
immediately after the initialisation phase. Consequently, the computational burden is larger than
would be required by the original algorithm to reach the desired precision specified by $r$ and
$s$ parameters in most cases. In order to reduce this unnecessary overhead, we consider the
interval $(\hat{\alpha}-\frac{\hat{\alpha}}{2},\hat{\alpha}+\frac{\hat{\alpha}}{2})$, which is
wider than the previous one whenever $\hat{\alpha} > \frac{1}{n_0}$ and leads to smaller sample
sizes.

The two-state Markov chain consists of two states $0$ and $1$, i.e., the two meta states of the
original DTMC. We set $\alpha$ as the probability of making the transition from state~$0$ to
state~$1$ (denoted as $0 \rightarrow 1$).
The estimate $\hat{\alpha}$ is computed as  the ratio of the number of transitions from state~$0$
to state~$1$ to the number of transition from state~$0$. Let $n_{0,\alpha}$ be the number of
transitions in the sample starting from state~$0$. Let $X_i$, $i = 1,2,\ldots,n_{0,\alpha}$, be
a~random variable defined as follows: $X_i$ is $1$ if $i$th transition from meta-state~$0$ is
$0\rightarrow 1$ and $0$ otherwise.

Notice that state~$0$ is an~accessible atom in the terminology of the theory of Markov chains,
i.e., the Markov chain regenerates after entering state~$0$, and hence the random variables
$X_i$, $i = 1,2,\ldots,n_{0,\alpha}$, are independent. They are Bernoulli distributed with
parameter $\alpha$. The unbiased estimate of the population variance from the sample, denoted
$\hat{\sigma}^2$, is given by
$\hat{\sigma}^2 = \hat{\alpha}\cdot(1-\hat{\alpha})\cdot\frac{n_{0,\alpha}}{n_{0,\alpha}-1}$.
Due to independence, $\hat{\sigma}^2$ is also the asymptotic variance and, in consequence, the
sample size that provides the specified confidence level for the estimate of the value of $\alpha$
is given by
$n_{\alpha,s}(\hat{\alpha},n_{0,\alpha}) =
\hat{\alpha}\cdot(1-\hat{\alpha})\cdot\frac{n_{0,\alpha}}{n_{0,\alpha}-1}\cdot
     \left(\frac{\mathrm{\Phi}^{-1}(\frac{1}{2}(1+s))}{\hat{\alpha}/2}\right)^2$.

The Markov chain is in state~$0$ with steady-state probability $\frac{\beta}{\alpha+\beta}$. Then,
given that the chain reached the steady-state distribution, the expected number of regenerations,
i.e., returns to meta state 0, in a~sample of size $n$ is given by
$\frac{n\cdot\beta}{\alpha+\beta}$. Therefore, the sample size used to estimate the value of
$\alpha$ with the specified confidence level $s$ is given by 
$n_{\alpha}=\frac{\alpha+\beta}{\beta}\cdot n_{\alpha,s}(\hat{\alpha},n_{0,\alpha})$. As the real
values of $\alpha$ and $\beta$ are unknown, the estimated values $\hat{\alpha}$ and $\hat{\beta}$
can be used in the above formula. If the computed $n_{\alpha}$ is bigger than the current number
of transitions $n_{0,\alpha}$, we extend the trajectory to reach $n_{\alpha}$ transitions from $0$
to $1$ and re-estimate the values for $\alpha$ and $\beta$ using the extended trajectory. We
repeat this process until the computed $n_{\alpha}$ value is smaller than the number of
transitions used to estimate $\alpha$. In this way, good initial estimates for $\alpha$ and
$\beta$ are obtained and the original two-state Markov chain approach using the formula for
$n(\alpha,\beta)$ is run.


\medskip
\noindent
{\bf Third approach: simple heuristics.}
When performing the initial estimation of $\alpha$ and $\beta$, we require both the count of
transitions from meta state~$0$ to meta state~$1$ and the count of transitions from meta-state~$1$
to meta state~$0$ be at least $3$. If this condition is not satisfied, we proceed by doubling the
length of the trajectory. In this way the problem of reaching the resolution boundary is
avoided. Our experiments showed that this simple approach in many cases led to good initial
estimates of the $\alpha$ and $\beta$ probabilities.

\section{Evaluation}
\label{sec:evaluation}
We implemented the two-state Markov chain approach with the simple heuristics presented in
Section~\ref{sec:twostate} and the Skart method of~\cite{TWLS08} in the tool {\sf ASSA-PBN},
which was specially designed for steady-state analysis of large PBNs~\cite{assa}
(see Section~\ref{ssec:pbn} for the theoretical background of PBNs).
We verified with experiments that with use of the simple heuristics,
the two-state Markov chain approach could meet the predefined precision requirement
even in the case of an unlucky initial sample size.
For the steady-state analysis of large PBNs, applications of these two
methods necessitate generation of trajectories of significant length. To achieve this in
an~efficient way, we applied the alias method~\cite{WAJ77} to sample the consecutive trajectory
state. This enables {\sf ASSA-PBN}, for example, to simulate $4,800$ steps within $1$s for
a~$2,000$ nodes PBN (state-space of size $2^{2,000}$).

%
We choose the Skart method~\cite{TWLS08} as a~reference for the evaluation of the performance
of the two-state Markov chain approach. The Skart method is a~successor of ASAP3, WASSP, and
SBatch methods, which are all based on the idea of batch means~\cite{TWLS08}. It is
a~procedure for on-the-fly statistical analysis of the simulation output, asymptotically
generated in accordance with a~steady-state distribution. Usually it requires an~initial sample of
size smaller than other established simulation analysis procedures~\cite{TWLS08}. In a~brief,
high-level summary, the algorithm partitions a~long simulation trajectory into batches, for each
batch computes a~mean and constructs an~interval estimate using the batch means. Further,
the interval estimate is used by Skart to decide whether a~steady state distribution is
reached or more samples are required. For a~more detailed description of this method,
see~\cite{TWLS08}.


The Skart method differs in two key points with the two-state Markov chain approach.
First, it specifies the initial trajectory length to be at least $1,280$, while for the two-state
Markov chain approach this information is not provided. 
Second, the Skart method applies the student distribution for skewness adjustment while the
two-state Markov chain approach makes use of the normal distribution for confidence interval
calculations.

To compare the performance of the two methods, we randomly generated $882$ different PBNs
using {\sf ASSA-PBN}. {\sf ASSA-PBN} can randomly generate a~PBN which satisfies structure
requirements given in the form of five input parameters: the node number, the minimum and the
maximum number of predictor functions per node, finally the minimum and maximum number of
parent nodes for a~predictor function. We generated PBNs with node numbers from the set
$\{15, 30, 80, 100, 150, 200, 300, 400, 500,\allowbreak 1000,\allowbreak 2000\}$. We assigned the
obtained PBNs into three different classes with respect to the density measure $\mathcal{D}$:
\emph{dense models} with density $150$--$300$, \emph{sparse models} with density around $10$, and
\emph{in-between models} with density $50$--$100$. The two-state Markov chain approach and the
Skart method were tested on these PBNs with precision $r$ set to the values in
$\{10^{-2},5\times10^{-3},10^{-3},5\times10^{-4},10^{-4},8\times10^{-5},5\times10^{-5}\}$. We set
$\epsilon$ to $10^{-10}$ for the two-state Markov chain approach and $s$ to $0.95$ for both
methods.

The experiments were performed on a~HPC cluster, with CPU speed ranging between $2.2$GHz and
$3.07$GHz. {\sf ASSA-PBN} is implemented in Java and the initial and maximum Java virtual machine
heap size were set to $503$MB and $7.86$GB, respectively. We collected $5263$ results with
the information on the PBN node number, its density class, the precision value, the estimated
steady-state probabilities computed by the two methods, and their CPU time costs. The steady-state
probabilities computed by the two methods are comparable in all the cases (data not shown in the
paper). For each experimental result $i$, we compare the time costs of the two methods. Let
$t_{\it TS}(i)$ and $t_{\it Skart}(i)$ be the time cost for the two-state Markov chain approach
and the Skart method, respectively. We say that the two-state Markov chain approach is by $k$ per
cent faster than the Skart method if $\frac{(t_{\it Skart}(i)-t_{\it TS}(i))}{t_{\it
Skart}(i)}\geqslant \frac{k}{100} $. The definition for the Skart method to be faster than the
two-state Markov chain approach is symmetric. In Table~\ref{tab:compare} we show the percentage of
cases in which the two-state Markov chain approach was by $k$ per cent faster than Skart and vice
versa for different $k$. In general, in about $70\%$ of the results, the two-state Markov chain
was faster than Skart. It is also clear that the number of cases the two-state Markov chain
approach was faster than Skart is larger than in the opposite case.

\begin{table}[!t]
  \centering
  \begin{tabular}{| c || r| r| r|  r| r| r| r|}
  \hline
  $k$ & 0 & 5 & 10 & 15 & 20 & 25 &30\\
  \hline\hline
  $t_{\it TS}\le t_{\it Skart}$ & \hspace{2mm}69.83\%	&\hspace{2mm}55.10\% &\hspace{2mm}41.02\%
  &\hspace{2mm}31.03\% &\hspace{2mm}25.86\% &\hspace{2mm}	22.76\% &\hspace{2mm}20.05\%\\
  \hline
  $t_{\it Skart}\le t_{\it TS}$ &30.38\% &18.68\%	&11.10\% &7.39\% &5.49\% &4.43\% &3.74\%\\
  \hline
  \end{tabular}
  \caption{Performance comparison of the Skart and the two-state MC methods. Explanations in the
  text.}\label{tab:compare}
\end{table}

We show in Table~\ref{tab:time} the trajectory sizes and the time costs for computing steady-state
probabilities of two large PBNs using the two-state Markov chain approach and the Skart method for
different precision requirements. The two analysed PBNs consist of $1,000$ and $2,000$ nodes,
which give rise to state spaces of sizes exceeding $10^{300}$ and $10^{600}$, respectively.
\begin{table}[!t]
\begin{tabular}{|p{1.1cm}||l|R{1.35cm}|R{1.35cm}|R{1.35cm}||R{1.35cm}|R{1.35cm}|R{1.35cm}|}
\hline
\multirow{2}{*}{\parbox{1.1cm}{node number}} & method      & \multicolumn{3}{c||}{the two-state Markov chain} & \multicolumn{3}{c|}{Skart} \\ \cline{2-8}
                             & precision       & 0.01    & 0.005   & 0.001   &   0.01    & 0.005   & 0.001       \\ \hline\hline
\multicolumn{1}{|r||}{\multirow{2}{*}{1,000}} 
                             & trajectory size &    35,066    &  133,803         &    3,402,637       &       37,999         &  139,672         &   3,272,940        \\ \cline{2-8}
                             & time cost  (s)     & 6.19&  23.53
       &    616.26     &  7.02       &   24.39      &    590.26           \\ \hline
\multicolumn{1}{|r||}{\multirow{2}{*}{2,000}} 
                             &    trajectory size             &     64,057      &    240,662      &   5,978,309 &63,674 &273,942  &5,936,060   \\ \cline{2-8}
                             &     time cost  (s)            &    20.42      &   67.60      &       1722.86&20.65 &78.53 &1761.05\\ \hline
\end{tabular}
\caption{Approximate steady-state analysis of two large PBNs.}
\label{tab:time}
\end{table}

\section{A Biological Case study}
\label{sec:casestudy}
A~multicellular organism consists of cells that form a~highly organised community. The number of
cells in this system is tightly controlled by mechanisms that regulate the cell division and the
cell death. One of these mechanisms is the programmed cell death, also referred to as \emph{apoptosis}: if cells are damaged, infected, or no longer needed, the intracellular death
program is activated, which leads to fragmentation of the DNA, shrinkage of the cytoplasm,
membrane changes and cell death without lysis or damage to neighbouring cells.
This process is regulated by a~number of signaling pathways which are extensively linked by
cross-talk interactions.
In~\cite{SSVSSBEMS09}, a~large-scale Boolean network of apoptosis in hepatocytes was introduced,
where the assigned Boolean interactions for each molecule were derived from literature study.
In~\cite{TMPTS14}, the original multi-value Boolean model was cast into the PBN framework:
a~binary PBN model, so-called \emph{`extended apoptosis model'} which comprised $91$ nodes
(state-space of size $2^{91}$) and $102$ interactions was constructed, see
Figure~\ref{fig:structure} in Appendix~\ref{app:apoptosis_model} for the wiring of the PBN model.
With respect to the original multi-value Boolean model of~\cite{SSVSSBEMS09}, the PBN model was
extended as described in~\cite{TMPTS14}. For example, the possibility of activation of
NF-$\kappa$B through Caspase~8 (\node{C8*}) was included. The model was fitted to steady-state
experimental data obtained in response to six different stimulations of the input nodes,
see~\cite{TMPTS14} for details.


As can be seen from the wiring of the network, the activation of \node{complex2} (\complex{2}) by
\node{RIP-deubi} can take place in two ways: 1)~by a~positive feedback loop from activated
\node{C8*} and \node{P} $\rightarrow$ \node{tBid} $\rightarrow$ \node{Bax} $\rightarrow$
\node{smac} $\rightarrow$ \node{RIP-deubi} $\rightarrow$ \complex{2} $\rightarrow$
\node{C8*-}\complex{2} $\rightarrow$ \node{C8*}, and 2)~by the positive signal from UV-B
irradiation (input nodes \node{UV(1)} or \node{UV(2)}) $\rightarrow$ \node{Bax} $\rightarrow$
\node{smac} $\rightarrow$ \node{RIP-deubi} $\rightarrow$ \complex{2}. The former to be active
requires the stimulation of the type 2 receptor (\node{T2R}). The latter way requires
\node{complex1} (\complex{1}) to be active, which cannot happen without the stimulation of the TNF
receptor-1. Therefore, \node{RIP-deubi} can activate \complex{2} only in the condition of
co-stimulation by \node{TNF} and either \node{UV(1)} or \node{UV(2)}. In consequence, it was
suggested in~\cite{TMPTS14} that the interaction of activation of \complex{2} via \node{RIP-deubi}
is not relevant and could be removed from the model in the context of modelling primary
hepatocyte. However, due to the problem with efficient generation of very long trajectories in
optPBN toolbox, quantitative analysis was hindered and this hypothesis could not be verified
(\cite{TMPTS14}).

In this work, we take up this challenge and we quantitatively investigate the relevancy of the
interaction of activation of \complex{2} via \node{RIP-deubi}. We perform an~extensive analysis in
the context of co-stimulation by \node{TNF} and either \node{UV(1)} or \node{UV(2)}: we compute
long-term influences of parent nodes on the \complex{2} node and the long-run sensitivities with
respect to various perturbations related to specific predictor functions and their selection
probabilities. For this purpose we apply the two-state Markov chain approach as implemented in our
{\sf ASSA-PBN} tool~\cite{assa} to compute the relevant steady-state probabilities for the
best-fit models described in~\cite{TMPTS14}. Due to the efficient implementation, the
{\sf ASSA-PBN} tool can easily deal with trajectories of length exceeding $2\times 10^9$ for this
case study.

We consider $20$ distinct parameter sets of~\cite{TMPTS14} that resulted in the best fit of the
`extended apoptosis model' to the steady-state experimental data in six different stimulation
conditions. In~\cite{TMPTS14}, parameter estimation was performed with steady-state measurements
for the nodes \node{apoptosis}, \node{C3ap17} or \node{C3ap17\_2} depending on the stimulation
condition considered, and \node{NF-$\kappa$B}. The optimisation procedure used was Particle Swarm
and fit score function considered was the sum of squared errors of prediction (SSE) and the sum
was taken over the three nodes in the six stimulation conditions. We took all the optimisation
results from the three independent parameter estimation runs of~\cite{TMPTS14}, each containing
$7500$ parameter sets. We sorted them increasingly with respect to the cost function value
obtained during optimisation, removed duplicates, and finally took the first $20$ best-fit
parameter sets.


As mentioned above, we fix the experimental context to co-stimulation of \node{TNF} and either
\node{UV(1)} or \node{UV(2)}. We note that originally in~\cite{SSVSSBEMS09} UV-B irradiation
conditions were imposed via a~multi-value input node \node{UV} which could take on three values,
i.e., 0 (no irradiation), 1 ($300\,J/m^2$ UV-B irradiation), and 2 ($600\,J/m^2$ UV-B
irradiation). In the model of~\cite{TMPTS14}, \node{UV} input node was refined as \node{UV(1)} and
\node{UV(2)} in order to cast the original model into the binary PBN framework. Therefore, we
consider in our study two cases: 1)~co-stimulation of \node{TNF} and \node{UV(1)} and
2)~co-stimulation of \node{TNF} and \node{UV(2)}.
Node \complex{2} has two independent predictor functions: \complex{2} = \complex{1} $\land$
\node{FADD} or \complex{2} = \complex{1} $\land$ \node{FADD} $\land$ \node{RIP-deubi}. The
selection probabilities are denoted as $c_1^{(\complex{2})}$ and $c_2^{(\complex{2})}$,
respectively. Their values have been optimised in~\cite{TMPTS14}.

We start with computing the influences with respect to the steady-state distribution, i.e.,
the long-term influences on \complex{2} of each of its parent nodes: \node{RIP-deubi},
\complex{1}, and \node{FADD}, in accordance with the definition in Section~\ref{ssec:pbn}.
Notice that the computation of the three influences requires several joint steady-state
probabilities to be estimated with the two-state Markov chain approach, e.g.,
(\complex{1}=1,\node{FADD}=1,\node{RIP-deubi}=0) or (\complex{1}=1,\node{FADD}=0). Each
probability determines a~specific split of the original Markov chain. For example, in the case
of the estimation of the joint steady-state probability for (\complex{1}=1,\node{FADD}=0), the
states of the underlying Markov chain of the apoptosis PBN model in which \complex{1}=1 and
\node{FADD}=0 constitute meta~state~1 and all the remaining states form meta~state~0. Therefore,
the estimation of influences is computationally demanding. The summarised results for the $20$
parameter sets are presented for the co-stimulation of \node{TNF} and \node{UV(1)} or \node{TNF}
and \node{UV(2)} in Table~\ref{tab:infl_complex2}. They are consistent across the different
parameter sets and clearly indicate that the influence of \node{RIP-deubi} on \complex{2} is small
compared to the influence of \complex{1} or \node{FADD} on \complex{2}. However, the influence of
\node{RIP-deubi} is not negligible.

\begin{table}[!t]
\centering
\begin{tabular}{|l|c|c|c||c|c|c|}
\cline{2-7}
\multicolumn{1}{c|}{} & \multicolumn{3}{c||}{\node{TNF} and \node{UV(1)}} &
\multicolumn{3}{c|}{\node{TNF} and \node{UV(2)}}\\
\hline
 & $I_\text{RIP-deubi}$ & $I_\text{\complex{1}}$ & $I_\text{FADD}$
 & $I_\text{RIP-deubi}$ & $I_\text{\complex{1}}$ & $I_\text{FADD}$\\
\hline\hline
Best fit & ~~~~0.2614 & ~~~~0.9981 & ~~~~0.9935 & ~~~~0.2615 & ~~~~0.9980 & ~~~~0.9936\\
\hline
Min & ~~~~0.0000 & ~~~~0.9979 & ~~~~0.9935 & ~~~~0.0000 & ~~~~0.9979 & ~~~~0.9936\\
\hline
Max & ~~~~0.3145 & ~~~~0.9988 & ~~~~0.9944 & ~~~~0.3146 & ~~~~0.9990 & ~~~~0.9947\\
\hline
Mean & ~~~~0.2087 & ~~~~0.9982 & ~~~~0.9937 & ~~~~0.2088 & ~~~~0.9982 & ~~~~0.9938\\
\hline
Std & ~~~~0.0735 & ~~~~0.0002 & ~~~~0.0002 & ~~~~0.0735 & ~~~~0.0002 & ~~~~0.0003\\
\hline
\end{tabular}
\caption{Long-term influences of \node{RIP-duebi}, \complex{1}, and \node{FADD} on \complex{2} in
the `extended apoptosis model' in~\cite{TMPTS14} under the co-stimulation of both \node{TNF} and
\node{UV(1)} or \node{UV(2)}.}
\label{tab:infl_complex2}
\end{table}

%

We take the analysis of the importance of the interaction between \node{RIP-deubi} and \complex{2}
further and we compute various long-run sensitivities with respect to selection probability
perturbation. In particular, we perturb the selection probability $c_2^{(\complex{2})}$ by $\pm
5\%$, i.e., we set the new value by multiplying the original value by $(1 \pm 0.05)$, and compute
in line with Definition~\ref{def:LRS_prob} how the joint steady-state distribution for
(\node{apoptosis},\node{C3ap17},\node{NF\-$\kappa$B}) differs from the non-perturbed one with
respect to the $l_1$~norm, i.e., $||\cdot||_1$. We notice that the computation of the full
steady-state distribution for the considered PBN model of apoptosis is practically intractable,
i.e., it would require the estimation of $2^{91}$ values. Therefore, we restrict the computations
to the estimation of eight joint steady-state probabilities for all possible combinations of
values for (\node{apoptosis},\node{C3ap17},\node{NF\-$\kappa$B}), i.e., the experimentally
measured nodes. Each estimation is obtained by a~separate run of the two-state Markov chain
approach with the split into meta states determined by the considered probability as explained
above in the case of the computation of long-term influences. To compare the estimated
distributions we choose the $l_1$~norm after~\cite{QD09}, where it is used in the computations of
similar types of sensitivities for PBNs to these defined in Section~\ref{ssec:pbn}. Notice that
the $l_1$~norm of the difference of two probability distributions on a~finite sample space is
twice the \emph{total variation distance}. The latter is a~well-established metric for measuring
the distance between probability distributions defined as the maximum difference between the
probabilities assigned to a~single event by the two distributions (see, e.g., \cite{LPW09}).
Additionally, we check the difference when $c_2^{(\complex{2})}$ is set to $0$ (and, in
consequence, $c_1^{(\complex{2})}$ is set to $1$). The obtained results for the $20$ parameter
sets in the conditions of co-stimulation of \node{TNF} and \node{UV(1)} and co-stimulation of
\node{TNF} and \node{UV(2)} are summarised in Table~\ref{tab:lrs_complex2}. In all these cases,
the sensitivities are very small. Therefore, the system turns to be insensitive to small
perturbations of the value of $c_2^{(\complex{2})}$. Also the complete removal of the second
predictor function for \complex{2} does not cause any drastic changes in the joint steady-state
distribution for (\node{apoptosis},\node{C3ap17},\node{NF-$\kappa$B}).

\begin{table}[!t]
\centering
\begin{tabular}{|l|c|c|c||c|c|c|}
\cline{2-7}
\multicolumn{1}{c|}{} & \multicolumn{3}{c||}{\node{TNF} and \node{UV(1)}} & \multicolumn{3}{c|}{\node{TNF} and \node{UV(2)}}\\
\hline
\hspace{2mm} $c_2^{(\complex{2})}$ \hspace{2mm} & \hspace{2mm} $+5\%$ \hspace{2mm} &
\hspace{2mm} $-5\%$ \hspace{2mm} & \hspace{4mm} $=0 \hspace{4mm}$ &
\hspace{2mm} $+5\%$ \hspace{2mm} & \hspace{2mm} $-5\%$ \hspace{2mm} &
\hspace{4mm} $=0 \hspace{4mm}$\\
\hline\hline
Best fit & 0.0003 & 0.0002 & 0.0011 & 0.0002 & 0.0004 & 0.0011\\
\hline
Min & 0.0002 & 0.0002 & 0.0003 & 0.0002 & 0.0002 & 0.0002\\
\hline
Max & 0.0008 & 0.0008 & 0.0014 & 0.0012 & 0.0007 & 0.0013\\
\hline
Mean & 0.0005 & 0.0005 & 0.0009 & 0.0004 & 0.0004 & 0.0009\\
\hline
Std & 0.0001 & 0.0001 & 0.0003 & 0.0002 & 0.0001 & 0.0003\\
\hline
\end{tabular}
\caption{Long-run sensitivities w.r.t\ selection probability perturbations for the `extended
apoptosis model' of~\cite{TMPTS14} under the co-stimulation of TNF and UV(1) or TNF and UV(2).}
\label{tab:lrs_complex2}
\end{table}


Finally, we compute the long-run sensitivity with respect to permanent on/off perturbations of the
node \node{RIP-deubi} in accordance with Definition~\ref{def:LRS_node}. As before, we consider the
joint steady-state distributions for (\node{apoptosis},\node{C3ap17},\node{NF-$\kappa$B}) and we
choose the $l_1$-norm. The results, given in Table~\ref{tab:lrs_RIP-deubi_pert}, show that in both
variants of UV-B irradiation the sensitivities are not negligible and the permanent on/off
perturbations of \node{RIP-deubi} have impact on the steady-state distribution.

\begin{table}[!t]
\centering
\begin{tabular}{|c||c|c|c|c|c|}
\hline
\hspace{1em} RIP-deubi f. pert. \hspace{1em} & \hspace{1em} Best fit \hspace{1em} &
\hspace{1em} Min \hspace{1em} & \hspace{1em} Max \hspace{1em} & \hspace{1em} Mean \hspace{1em} &
\hspace{1em} Std \hspace{1em} \\
\hline\hline
\node{TNF} \& \node{UV(1)} & 0.3075 & 0.0130 & 0.3595 & 0.2089 & 0.0823 \\
\hline
\node{TNF} \& \node{UV(2)} & 0.3097 & 0.0105 & 0.3612 & 0.2105 & 0.0827 \\
\hline
\end{tabular}
\caption{Long-run sensitivities w.r.t\ permanent on/off perturbations of \node{RIP-deubi} for the
`extended apoptosis model' of~\cite{TMPTS14}.}
\label{tab:lrs_RIP-deubi_pert}
\end{table}

To conclude, all the obtained results indicate that in the context of co-stimulation of \node{TNF}
and either \node{UV(1)} or \node{UV(2)} the interaction between \node{RIP-deubi} and \complex{2}
plays a~certain role. Although the elimination of the interaction does not invoke significant
changes to the considered joint steady-state distribution, the long-term influence of
\node{RIP-deubi} on \complex{2} is not negligible and may be important for other nodes in the
network other than \node{apoptosis}, node{C3ap17}, or \node{NF-$\kappa$B}.

\section{Discussion and Conclusion}
\label{sec:conclusion}

In this paper, we focused on two statistical methods for estimating steady-state probabilities of
large PBNs: the two-state Markov chain approach and the Skart method. The Skart method follows
a~continuous development~\cite{TWLS08}, while the two-state Markov chain approach was originally
introduced by Raftery and Lewis in 1992, and only recently it was explored for the analysis of
a~relatively large PBN model in~\cite{TMPTS14}. To revive the application of the two-state Markov
chain approach, we propose a~few heuristics to avoid a~problem with the size of the initial
sample which can lead to biased results. By extensive experiments, we show that the two-state
Markov chain approach outperforms the Skart method in most cases. In the end, we illustrated the
usability of the two-state Markov chain approach on a~realistic biological system.

Our work in the current paper is closely related to statistical model checking~\cite{YS02,SVA05},
a~simulation-based approach using hypothesis testing to infer whether a~stochastic system
satisfies a~property. Most current tools for statistical model checking are restricted for bounded
properties which can be checked on finite executions of the system. In recent year, both the Skart
method and the perfect simulation algorithm have been explored for statistical model checking of
steady state and unbounded until properties~\cite{RP09,Roh13}, which was considered as a~future
step of statistical model checking~\cite{LDB10}. The perfect simulation algorithm for sampling the
steady-state of an ergodic~DTMC is based on the indigenous idea of the \emph{backward coupling
scheme} originally proposed by Propp and Wilson in~\cite{PW96}. It allows to draw independent
samples which are distributed exactly in accordance with the steady-state distribution of a~DTMC.
However, due to the nature of this method, each state in the state space needs to be considered at
each step of the coupling scheme. Of course, a~special, more efficient variant of this method
exists. If a~DTMC is monotone, then it is possible to sample from the steady-state distribution
by considering the maximal and minimal states only~\cite{PW96,BGV08}. For example, this approach
was exploited in~\cite{RP09} for model checking large queuing networks. Unfortunately, it is not
applicable in the case of PBNs with perturbations. In consequence, the perfect simulation
algorithm is only suited for at most medium-size PBNs and large-size PBNs are out of its scope.
Thus, in this paper we have only compared the performance of the two-state Markov chain approach
with the Skart method.

Moreover, in this study we have identified a~problem of generating biased results by the original
two-state Markov chain approach and have proposed three heuristics to avoid wrong initialisation.
Finally, we demonstrated the potential of the two-state Markov chain approach on a~study of
a~large, $91$-node PBN model of apoptosis in hepatocytes. The two-state Markov chain approach
facilitated the quantitative analysis of the large network and the investigation of a~previously
formulated hypothesis regarding the relevance of the interaction of activation of \complex{2} via
\node{RIP-deubi}. In the future, we aim to investigate the usage of the discussed statistical
methods for approximate steady-state analysis in a~wide project on systems biology. For instance,
we will further apply them to develop new techniques for minimal structural interventions to alter
steady-state probabilities, which will enable the synthesis of optimal control strategies for
large regulatory networks.

\medskip
\noindent
\textbf{Acknowledgment.}
Experiments presented in this paper were carried out using the HPC facilities of the University of Luxembourg~\cite{VBCG_HPCS14}
(\url{http://hpc.uni.lu}).

\bibliographystyle{splncs}
\bibliography{twoState}

\newpage


\appendix

\section{Derivation of the number of ``burn-in'' iterations}
\label{app:m}
Let $\{Z_t\}_{t\geq 0}$ be a~discrete-time two-state Markov chain as given in
Figure~\ref{fig:two-state_MC}.
$Z_t$ has the value $0$ or $1$ if the system is in state $0$ or state $1$ at time $n$,
respectively. The transition probabilities satisfy $0<\alpha,\beta<1$ and the transition matrix
for this chain has the following form
\[
P = \left[ \begin{array}{cc}
1-\alpha & \alpha \\
\beta & 1-\beta
\end{array} \right].
\]
Matrix $P$ has two distinct eigenvalues: $1$ and $\lambda = (1-\alpha-\beta)$. Notice that
$|\lambda| < 1$.

The chain is ergodic and the unique steady-state distribution is $\pi = [\pi_0\ \pi_1] =
[\frac{\beta}{\alpha+\beta}\ \frac{\alpha}{\alpha+\beta}]$.
Let $\mathbb{E}_\pi(Z_t)$ denote the expected value of $Z_t$ for any fixed $t\geq 0$, with respect
to the steady-state distribution $\pi$. We have that
$\mathbb{E}_\pi(Z_t)=\frac{\alpha}{\alpha+\beta}$.

The $m$-step transition matrix can be written, as can be checked by induction, in the form
\begin{equation}
\label{eq:P}
P^m =
\left[ \begin{array}{cc}
\pi_0 & \pi_1 \\
\pi_0 & \pi_1
\end{array} \right] +
\frac{\lambda^m}{\alpha+\beta}\cdot
\left[ \begin{array}{cc}
\alpha & -\alpha \\
-\beta & \beta
\end{array} \right],
\end{equation}
where $\lambda$ is the second eigenvalue of $P$.

Suppose we require $m$ to be such that the following condition is satisfied
\begin{equation}
\label{eq:m-condition}
\Big|\left[ \begin{array}{c}
\mathbb{P}[Z_m=0\,|\,Z_0=j] ~~
\mathbb{P}[Z_m=1\,|\,Z_0=j]
\end{array} \right]
-
\left[ \begin{array}{c}
\pi_0 ~~
\pi_1
\end{array} \right]
\Big|
<
\left[ \begin{array}{c}
\epsilon ~~
\epsilon
\end{array} \right]
\end{equation}
for some $\epsilon > 0$. For any vector $v=[v_1\ v_2\ \ldots\ v_n]^T \in\mathbb{R}^n$ we use $|v|$
to denote $[|v_1|\ |v_2|\ \allowbreak \ldots\ \allowbreak |v_n|]^T$, where $T$ is the
transposition operator. If $e_0=[1\ 0]$ and $e_1 = [0\ 1]$, then for $j \in \{0,1\}$ we have that
\begin{equation}
\label{eq:m-distribution}
\left[ \begin{array}{c}
\mathbb{P}[Z_m=0\,|\,Z_0=j] ~~
\mathbb{P}[Z_m=1\,|\,Z_0=j]
\end{array} \right]
=e_jP^m.
\end{equation}
With \eqref{eq:P} and \eqref{eq:m-distribution}, condition \eqref{eq:m-condition} can be rewritten
as
\[
\Bigg|e_j\left(
\left[ \begin{array}{cc}
\pi_0 & \pi_1 \\
\pi_0 & \pi_1
\end{array} \right] +
\frac{\lambda^m}{\alpha+\beta}\cdot
\left[ \begin{array}{cc}
\alpha & -\alpha \\
-\beta & \beta
\end{array} \right]
\right)
-
\left[ \begin{array}{c}
\pi_0 ~~
\pi_1
\end{array} \right]
\Bigg| <
\left[ \begin{array}{c}
\epsilon ~~
\epsilon
\end{array} \right].
\]
For $j = 0$ and $j = 1$ the above simplifies to
\[
\Bigg|
\frac{\lambda^m}{\alpha+\beta}\cdot
\left[ \begin{array}{c}
\alpha ~~
-\alpha
\end{array} \right]
\Bigg| <
\left[ \begin{array}{c}
\epsilon ~~
\epsilon
\end{array} \right]
\qquad \textrm{and} \qquad
\Bigg|
\frac{\lambda^m}{\alpha+\beta}\cdot
\left[ \begin{array}{c}
-\beta ~~
\beta
\end{array} \right]
\Bigg| <
\left[ \begin{array}{c}
\epsilon ~~
\epsilon
\end{array} \right],
\]
respectively. Therefore, it is enough to consider the following two inequalities
\[
\left|\frac{\lambda^m\alpha}{\alpha+\beta}\right|<\epsilon
\qquad \textrm{and} \qquad
\left|\frac{\lambda^m\beta}{\alpha+\beta}\right|<\epsilon,
\]
which, since $\alpha,\beta > 0$, can be rewritten as
\[
\left|\lambda^m\right|<\frac{\epsilon(\alpha+\beta)}{\alpha}
\qquad \textrm{and} \qquad
\left|\lambda^m\right|<\frac{\epsilon(\alpha+\beta)}{\beta}.
\]
Equivalently, $m$ has to satisfy
\[
\left|\lambda^m\right|<\frac{\epsilon(\alpha+\beta)}{\max(\alpha,\beta)}.
\]
By the fact that $|\lambda^m|=|\lambda|^m$ this can be expressed as
\[
|\lambda|^m<\frac{\epsilon(\alpha+\beta)}{\max(\alpha,\beta)}.
\]
Then, by taking the logarithm to base $10$ on both sides\footnote{In fact, by the formula for
change of base for logarithms, the natural logarithm ($\ln$), the logarithm to base $2$
($\log_2$), or a~logarithm to any other base could be used to calculate $m$ instead of $\log$.
Notice that $m$ does \textbf{not} depend on the choice of the base of the logarithm!}, we have
that
\[
m\cdot\log{(|\lambda|)}<\log{\left(\frac{\epsilon(\alpha+\beta)}{\max(\alpha,\beta)}\right)}
\]
and in consequence, since $|\lambda|<1$ and $\log{|\lambda|} < 0$,
\[
m>\frac{\log{\left(\frac{\epsilon(\alpha+\beta)}{\max(\alpha,\beta)}\right)}}{\log{(|\lambda|)}}.
\]

\section{Derivation of the sample size}
\label{app:n}
By the Law of Large Numbers for irreducible positive recurrent Markov chains
$\overline{Z}_n \to \pi_1\ a.\ s.$ with $n \to \infty$, where $\overline{Z}_n =
\frac{1}{n}\sum_{t=1}^nZ_t$.
%
%
Now, by a~variant of the Central Limit Theorem for non-independent random
variables\footnote{Notice that the random variables $Z_t$, $Z_{t+1}$ which values are consecutive
states of a~trajectory are correlated and are not independent.}, for $n$ large, $\overline{Z}_n$
is approximately normally distributed with mean $\pi_1=\frac{\alpha}{\alpha+\beta}$ and asymptotic
variance
$\sigma_{\textrm{as}}^2 = \frac{1}{n}\frac{\alpha\beta(2-\alpha-\beta)}{(\alpha+\beta)^3}$, see
Section~\ref{app:as_var} for the derivation of the asymptotic variance. Let $X$ be the
standardised $\overline{Z}_n$, i.e.,
\[
X=\frac{\overline{Z}_n-\pi_1}{\sigma_{\textrm{as}}/\sqrt{n}}.
\]
If follows that $X$ is normally distributed with mean $0$ and variance $1$, i.e., $X \sim N(0,1)$.

Now, we require $n$ to be such that the condition $\mathbb{P}[\pi_1-r\leq \overline{Z}_n \leq
\pi_1+r] = s$ is satisfied for some specified $r$ and $s$. This condition can be rewritten as
\[
\mathbb{P}[-r\leq \overline{Z}_n-\pi_1 \leq r] = s,
\]
and further as
\[
\mathbb{P}[-r\cdot\frac{\sqrt{n}}{\sigma_{\textrm{as}}}\leq
 \frac{\overline{Z}_n-\pi_1}{\sigma_{\textrm{as}}/\sqrt{n}}
 \leq r\cdot\frac{\sqrt{n}}{\sigma_{\textrm{as}}}] = s,
\]
which is
\[
\mathbb{P}[-r\cdot\frac{\sqrt{n}}{\sigma_{\textrm{as}}}\leq X \leq
 r\cdot\frac{\sqrt{n}}{\sigma_{\textrm{as}}}] = s.
\]
Since $X \sim N(0,1)$ and $N(0,1)$ is symmetric around $0$, it follows that
\[
\mathbb{P}[0 \leq X \leq r\cdot\frac{\sqrt{n}}{\sigma_{\textrm{as}}}] = \frac{s}{2}
\]
and
\[
\mathbb{P}[X \leq r\cdot\frac{\sqrt{n}}{\sigma_{\textrm{as}}}] =
\frac{1}{2}+\frac{s}{2}=\frac{1}{2}(1+s).
\]
Let $\Phi(\cdot)$ be the standard normal cumulative distribution function. Then the above
can be rewritten as
\[
\Phi(r\cdot\frac{\sqrt{n}}{\sigma_{\textrm{as}}}) = \frac{1}{2}(1+s).
\]
Therefore, if we denote the inverse of the standard normal cumulative distribution function
with $\Phi^{-1}(\cdot)$, we have that
\[
r\cdot\frac{\sqrt{n}}{\sigma_{\textrm{as}}} = \Phi^{-1}(\frac{1}{2}(1+s)).
\]
In consequence,
\[
n = \frac{\sigma_{\textrm{as}}^2}{\left\{\frac{r}{\Phi^{-1}(\frac{1}{2}(1+s))}\right\}^2}=
\frac{\frac{\alpha\beta(2-\alpha-\beta)}{(\alpha+\beta)^3}}
{\left\{\frac{r}{\Phi^{-1}(\frac{1}{2}(1+s))}\right\}^2}.
\]

\section{Derivation of the asymptotic variance}
\label{app:as_var}

By the Central Limit Theorem for stationary stochastic processes\footnote{After discarding the
`burn-in' part of the trajectory, we can assume that the Markov chain in a~stationary stochastic
process.} $\sqrt{n}(\overline{Z}_n-\pi_1)\xrightarrow{d}N(0,\sigma_{\textrm{as}}^2)$ as
$n\to\infty$, where $\sigma_{\textrm{as}}^2$ is the so-called asymptotic variance given by
\begin{equation}
\label{eq:clt_var}
\sigma_{\textrm{as}}^2 = \Var_\pi(Z_j) + 2\sum_{k=1}^\infty \Cov_\pi(Z_j,Z_{j+k})
\end{equation}
and $\Var_\pi(\cdot)$ and $\Cov_\pi(\cdot)$ denote the variance and covariance with respect to the
steady-state distribution $\pi$, respectively. We proceed to calculate $\sigma_{\textrm{as}}^2$.
First, observe that $\E_\pi(Z_n\cdot Z_{n+1}) = \frac{\alpha}{\alpha+\beta}(1-\beta)$: $Z_n\cdot
Z_{n+1} \neq 0$ if and only if the chain is state $1$ at time $n$ and remains in $1$ at time
$n+1$, i.e., $Z_n = Z_{n+1} = 1$. The probability of this event at steady state is
$\frac{\alpha}{\alpha+\beta}(1-\beta)$. Then, by the definition of covariance, we have that the
steady-state covariance between consecutive random variables of the two-state Markov chain, i.e., $\Cov_\pi(Z_{n},Z_{n+1})$ is
\begin{align*}
\Cov_\pi(Z_{n},Z_{n+1})&=\E_\pi\left[(Z_{n}-\E_\pi(Z_{n}))(Z_{n+1}-\E_\pi(Z_{n+1}))\right]\\
&=\E_\pi\left[(Z_n-\frac{\alpha}{\alpha+\beta})(Z_{n+1}-\frac{\alpha}{\alpha+\beta})\right]\\
&=\E_\pi\left[Z_nZ_{n+1}-\frac{\alpha}{\alpha+\beta}(Z_n+Z_{n+1})+
  \frac{\alpha^2}{(\alpha+\beta)^2}\right]\\
&=\E_\pi(Z_nZ_{n+1})-\frac{\alpha}{\alpha+\beta}(\E_\pi(Z_n)+\E_\pi(Z_{n+1}))
  +\frac{\alpha^2}{(\alpha+\beta)^2}\\
&=\frac{\alpha(1-\beta)}{\alpha+\beta}-2\frac{\alpha^2}{(\alpha+\beta)^2}
  +\frac{\alpha^2}{(\alpha+\beta)^2}\\
&=\frac{\alpha\beta(1-\alpha-\beta)}{(\alpha+\beta)^2}.
\end{align*}
Further, we have that
$\Var_\pi(Z_n)=\pi_{0} \cdot \pi_1=\frac{\alpha\beta}{(\alpha+\beta)^2}$ (variance of the
Bernoulli distribution) and it can be shown that
$\Cov_\pi(Z_{n},Z_{n+k})=(1-\alpha-\beta)^k\cdot \Var_\pi(Z_{n})$ for any $k\geq 1$. Now,
according to Equation~\eqref{eq:clt_var}, we have
\begin{align*}
\sigma_{\textrm{as}}^2&=\Var_\pi(X_j) + 2\sum_{k=1}^\infty \Cov_\pi(X_j,X_{j+k})\\
&=\frac{\alpha\beta}{(\alpha+\beta)^2}+2\sum_{k=1}^\infty(1-\alpha-\beta)^k
  \cdot\frac{\alpha\beta}{(\alpha+\beta)^2}\\
&=\frac{\alpha\beta}{(\alpha+\beta)^2}+\frac{2\alpha\beta}{(\alpha+\beta)^2}
  \cdot\sum_{k=1}^\infty(1-\alpha-\beta)^k\\
&=\frac{\alpha\beta}{(\alpha+\beta)^2}+\frac{2\alpha\beta}{(\alpha+\beta)^2}
  \cdot\frac{1-\alpha-\beta}{\alpha+\beta}\\
&=\frac{\alpha\beta(2-\alpha-\beta)}{(\alpha+\beta)^3}.
\end{align*}
In consequence, $\overline{Z}_n$ is approximately normally distributed with mean
$\frac{\alpha}{\alpha+\beta}$ and variance
$\frac{1}{n}\frac{\alpha\beta(2-\alpha-\beta)}{(\alpha+\beta)^3}$.

\section{Derivations for the pitfall avoidance heuristics}
\label{app:1st_approach}

We start with analysing the minimum values $n(\cdot,\cdot)$ can attain. The function is considered
on the domain $D=(0,1]\times(0,1]$ and, as mentioned before, the estimated values of $\alpha$ and
$\beta$ are within the range $[\frac{1}{n_0},1]$. Computing the partial derivatives, equating them
to zero, and solving for $\alpha$ and $\beta$ yields $\alpha = -\beta$, which has no solution in
the considered domain. Hence, the function has neither local minimum nor maximum on $D$. Let us
fix $\beta$ for a~moment and consider $n(\alpha,\beta)$ as a~function of $\alpha$.
We denote it as $n_{\beta}(\alpha)$.
By differentiating with respect to $\alpha$, we obtain
\[
\frac{\partial}{\partial \alpha}n_{\beta}\left(\alpha\right) =
\frac{1}{c_{r,s}}\frac {\beta\, \left( {\alpha}^{2}-{\beta}^{2}-4\,\alpha+2\,\beta
 \right) }{ \left( \alpha+\beta \right) ^{4}}, \mbox{~where~}
c_{r,s} = \frac{r^2}{\left(\mathrm{\Phi}^{-1}\left(\frac{1}{2}(1+s)\right)\right)^2}.
\]
By equating to zero and solving for $\alpha$ we get two solutions:
$\alpha_1 = 2-\sqrt{\beta^2-2\beta+4}$ and
$\alpha_2 = 2+\sqrt{\beta^2-2\beta+4}$.
Since the second solution is always greater than $1$ on the $(0,1]$ interval, only the first
solution is valid. The sign of the second derivative of $n_{\beta}(\alpha)$ with respect to $\alpha$
at $\alpha_1$ is negative. This shows that for any fixed $\beta$, $n_{\beta}(\alpha)$ grows on the
interval $[\frac{1}{n_0},\alpha_1]$, attains its maximum at $\alpha_1$ and decreases on the
interval $[\alpha_1,1]$. Notice that $n$ is symmetric, i.e., $n(\alpha,\beta) = n(\beta,\alpha)$.
Thus the minimum value $n$ could attain for $\alpha$ and $\beta$ estimated from
a~sample of size $n_0$ is given by
$\min\left(n\left(\frac{1}{n_0},\frac{1}{n_0}\right),n\left(\frac{1}{n_0},1\right)\right)$.
After evaluating $n$ we get
\[
n\left(\frac{1}{n_0},\frac{1}{n_0}\right) = \frac{n_0-1}{4\,c_{r,s}}
\qquad\textrm{and}\qquad
n\left(\frac{1}{n_0},1\right) = \frac{(n_0-1)\cdot n_0}{c_{r,s}\cdot(1+n_0)^3}.
\]
Now, to avoid the situation where the initial estimates of $\alpha$ and $\beta$ lead to
$n(\alpha,\beta)<2\,n_0$, it is enough to make sure that given $r$ and $s$ the following condition
is satisfied:
$\min(n(\frac{1}{n_0},\frac{1}{n_0}),n(\frac{1}{n_0},1)) \geqslant 2\,n_0$.
This can be rewritten as
\begin{equation*}
\left\{\begin{array}{l}
(8\,c_{r,s}-1)\,n_0 + 1 \leq 0\\\\
2\,c_{r,s}\,n_0^3+6\,c_{r,s}\,n_0^2+(6\,c_{r,s}-1)\,n_0+2\,c_{r,s}+1 \leq 0
\end{array}
\right.
\end{equation*}
Both inequalities can be solved analytically. Given that $n_0>0$, the solution of the first
inequality is
\begin{equation}
\left\{\begin{array}{lp{1cm}l}
n_0 \in [-\frac{1}{8\cdot c_{r,s}-1},\infty) & & c_{r,s} < \frac{1}{8}\\
n_0 \in \emptyset & & c_{r,s} \geqslant \frac{1}{8}.
\end{array}
\right.
\end{equation}
The solution of the second inequality is more complicated, but can be easily obtained with
computer algebra system software (e.g., Maple\textsuperscript{\tiny TM}). In Table~\ref{tab:n_0}
we present some solutions for a~number of values for $r$ and $s$.

\newpage

\section{The Boolean model of apoptosis}
\label{app:apoptosis_model}

\begin{figure}[!h]
\centering
\includegraphics[height=.68\textwidth,angle=90]{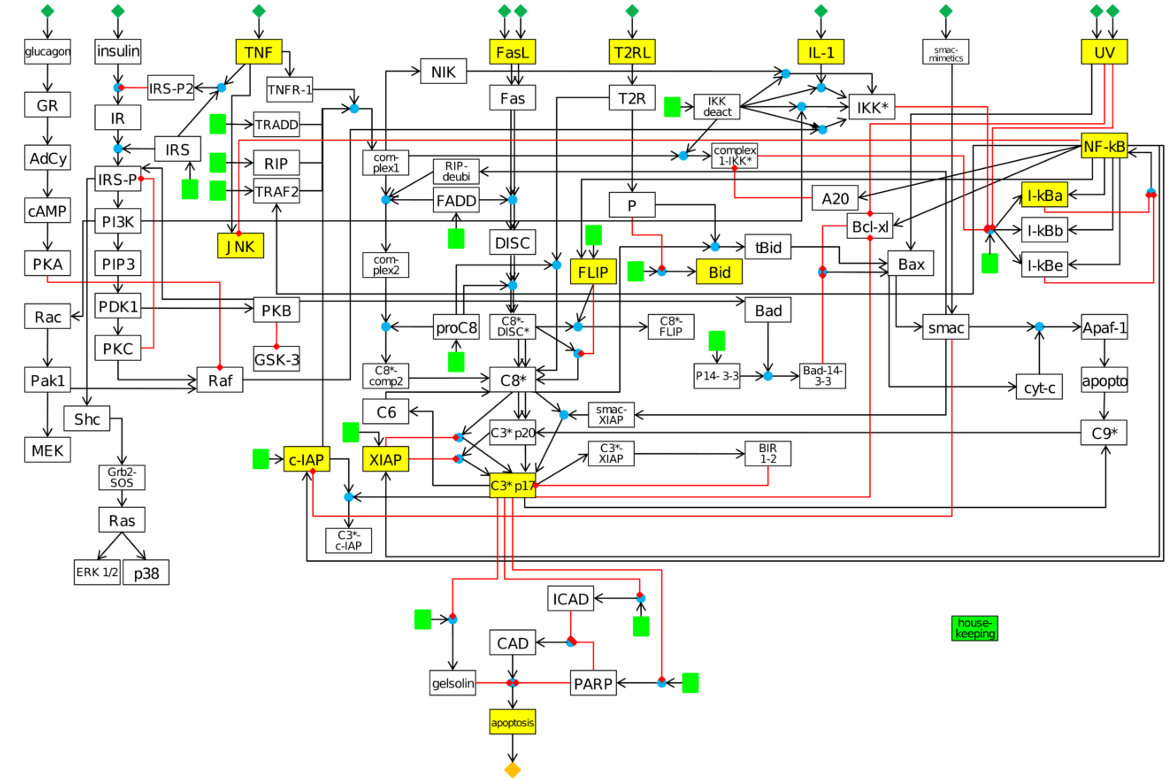}
\caption{The wiring of the probabilistic Boolean model of apoptosis originally introduced
in~\cite{TMPTS14}.}
\label{fig:structure}
\end{figure}

\end{document}